\documentclass[twocolumn,showpacs,aps,prl,amsmath,amssymb,nofootinbib,nobibnotes,floatfix]{revtex4-1}

\usepackage{hyperref}
\usepackage{amsfonts}
\usepackage{amsmath}
\usepackage{mathrsfs}
\usepackage{epsfig}
\usepackage{graphicx}               
\usepackage{url}
\usepackage{hyperref}
\usepackage{float}
\usepackage{color}
\usepackage[utf8]{inputenc}

\usepackage{graphicx}
\usepackage{epsf}

\newcommand{\ssec}[1]{{{\em #1.}---}}
\usepackage{comment}

\newcommand{\nc}{\newcommand}

\nc{\beq}{\begin{equation}}
\nc{\eeq}{\end{equation}}
\nc{\beqa}{\begin{eqnarray}}
\nc{\eeqa}{\end{eqnarray}}

\usepackage{slashed}

\newcommand{\lsim}{\!\mathrel{\hbox{\rlap{\lower.55ex \hbox{$\sim$}} \kern-.34em \raise.4ex \hbox{$<$}}}}
\newcommand{\gsim}{\!\mathrel{\hbox{\rlap{\lower.55ex \hbox{$\sim$}} \kern-.34em \raise.4ex \hbox{$>$}}}}

\def\be{\begin{equation}}
\def\ee{\end{equation}}

\newcommand\affspc{\vspace{4pt}}

\usepackage{footmisc}
\usepackage{setspace}
\begin{document}

\title{Massive Boson Superradiant Instability of Black Holes: Nonlinear Growth, Saturation, and Gravitational Radiation}

\author{William E.\ East}
\affiliation{Perimeter Institute for Theoretical Physics, Waterloo, Ontario N2L 2Y5, Canada \affspc}

\begin{abstract}
We study the superradiant instability of a massive boson around a spinning
black hole in full general relativity without assuming spatial
symmetries. We focus on the case of a rapidly spinning black hole in the presence of a 
vector boson with a Compton wavelength comparable to the black hole radius,
which is the regime where relativistic effects are maximized.  We follow
the growth of the boson cloud through superradiance into the nonlinear
regime as it spins down the black hole, reaches a maximum energy, and begins
to dissipate through the emission of gravitational waves.  
We find that the
superradiant instability can efficiently convert a significant fraction
of a black hole's rotational energy into gravitational radiation.
\end{abstract}

\maketitle

\ssec{Introduction}%
The advent of gravitational wave (GW) astronomy~\cite{Abbott:2016blz} not only
provides a window into the dynamics of merging black hole (BH) and neutron star
binaries, but also furnishes a tool to look for new fundamental physics.  One
example of beyond standard model particle physics that could be explored is
ultralight bosons that are weakly coupled to ordinary matter, including the QCD
axion~\cite{Weinberg:1977ma}, the string
axiverse~\cite{Arvanitaki:2009fg,Arvanitaki:2010sy}, and dark
photons~\cite{Holdom:1985ag,Cicoli:2011yh}.  Massive bosons are unstable in the presence of
spinning BHs, and can form clouds that grow exponentially through
superradiance~\cite{Damour:1976,Detweiler:1980uk,Zouros:1979iw}, potentially 
reaching masses up to $\sim10\%$ of the BH, and spinning the BH down in the
process~\cite{Arvanitaki:2010sy,Dolan:2012yt,2015CQGra..32m4001B,East:2017ovw}.

Thus, the existence of such bosonic fields can be probed indirectly by measuring the
masses and spins of BHs through electromagnetic observations of accretion disks
or GW observations of BH
mergers~\cite{Arvanitaki:2014wva,Arvanitaki:2016qwi,Baryakhtar:2017ngi,Baumann:2018vus}.  One
could also hope to directly observe the GWs coming from these oscillating
bosonic clouds, either from resolved
sources~\cite{Arvanitaki:2014wva,Arvanitaki:2016qwi,Baryakhtar:2017ngi}, or
from a stochastic background~\cite{Brito:2017zvb,Brito:2017wnc} with LIGO, LISA, or
other GW detectors. 

The goal of this work is to study the nonlinear growth and saturation of the
superradiant instability of massive bosons in the regime where relativistic
effects are maximized, both to understand the dynamics and GW signature in this
regime, and to place an upper limit on the importance of
nonlinear or ``nonadiabatic" effects.  To this end, we focus on the case of
a massive vector field, which has a significantly faster growth rate and stronger gravitational
radiation~\cite{East:2017mrj} than the scalar case, and consider its
superradiant growth around a rapidly spinning BH with size comparable to the
boson Compton wavelength.

There has been much recent progress in understanding the linear regime of the
superradiant instability of massive vector fields
~\cite{Pani:2012bp,Pani:2012vp,Endlich:2016jgc,Witek:2012tr,Baryakhtar:2017ngi,East:2017mrj,Cardoso:2018tly,Frolov:2018ezx,Dolan:2018dqv}.
There have, however, only been a handful of investigations of the superradiance of
rotating BHs beyond the test field limit, including the superradiant scattering
of a large amplitude GW~\cite{East:2013mfa}, and the nonlinear growth of the
superradiant instability in anti-de Sitter~\cite{Chesler:2018txn}.
In addition, there have been several studies of the analogous process for
charged BHs in spherical
symmetry~\cite{Sanchis-Gual:2015lje,Bosch:2016vcp,Baake:2016oku}

In Ref.~\cite{East:2017ovw}, the growth of the superradiant instability of a massive
vector field was studied, and found to be efficient even into the nonlinear
regime, with the instability smoothly saturating as the BH horizon frequency
decreases to match that of the oscillating cloud (see
also Ref.~\cite{Zilhao:2015tya} for a nonlinear study around a nonspinning BH). That
study focused on a complex vector field with an assumed $m=1$ azimuthal
symmetry that gave rise to an axisymmetric spacetime. This assumption reduces
the computational expense of a problem that is challenging due to the large
separation of timescales between the oscillation period and growth time of the
superradiant instability.  However, in such a setting the GW radiation and
other nonaxisymmetric couplings that would arise through gravity are
suppressed. There  also exist hairy BH
solutions~\cite{Herdeiro:2016tmi,Herdeiro:2017phl} as the plausible end point of
the instability fixing this symmetry~\cite{Ganchev:2017uuo,Degollado:2018ypf}, which is not the case for a real
vector field. 

Here we tackle for the first time the more challenging---and observationally
interesting---problem of a real massive vector field with no assumed spatial
symmetries, leveraging accurate initial data for an exponentially growing boson
cloud.  We show evidence of nonlinear mode coupling through gravity, both in
the GWs and in the field. 
However, we find that even in this regime where the growth
rate and gravitational radiation reaction are maximized, 
and where the saturation of the instability leads to the formation of 
a cloud rapidly oscillating close to the BH horizon with $\approx 6\%$ of its mass,
nonlinear effects do not significantly hamper the extraction, and conversion
into GWs, of the BH's rotational energy.

\ssec{Methodology}%
To study the superradiant instability, we evolve 
the Einstein field equations coupled to a (real) massive vector 
field $X^a$---also known as a Proca field---with mass parameter $\mu$ (boson mass divided by $\hbar$),
governed by
$\nabla_a F^{ab} = \mu^2 X^b$, 
with $F_{ab} = \nabla_a X_b-\nabla_b X_a$.
We evolve these equations using the same methods as described in Ref.~\cite{East:2017ovw},
except that our domain is fully three dimensional and does not assume any spatial
symmetries.
We use units with $G=c=1$ throughout.

In order to make it numerically tractable to follow the growth of the
superradiant instability through saturation, we need to start with consistent
initial data describing a non-negligible Proca cloud arising from the
superradiant instability.  To do that, we leverage the results
of Ref.~\cite{East:2017ovw} where a \emph{complex} massive vector field was evolved
on an axisymmetric spacetime. We focus on a case from there with $\mu=0.4/M_0$
around a BH with initial mass $M_0$ and dimensionless spin
$a=0.99$\footnote{This choice of parameters gives both a growth rate for the
linear instability, and an expected saturation energy for the field, that are
within a factor of two of the maximum values.
} and use the metric and Proca field
configurations after the Proca cloud has grown from $\sim10^{-3}M_0$ to
$10^{-2}M_0$, which is $\sim 1/6$ its saturation value. We then take just the
real part of the Proca field rescaled to give approximately the same total mass, i.e.
$X_a \rightarrow \sqrt{2} {\rm Re} ( X_a )$, and use this for constructing initial
data. There will be a small correction to the metric since we have a
transformed an axisymmetric stress energy into a nonaxisymmetric one.  We
therefore obtain consistent initial data by solving the Einstein constraint
equations using the methods of Ref.~\cite{East:2013mfa}, and using the above
configurations as the free data.  This slightly decreases the BH mass and
angular momentum, by $\sim0.3\%$, compared to the axisymmetric spacetime.  We
give more details on this procedure in the appendix.

In order to evolve these systems for timescales of $>10^4M_0$ at modest
resolution, without accumulating large numerical errors just from the isolated
BH spacetime, we use the background error subtraction technique~\cite{best}.
Unless otherwise noted, the results presented were obtained using a numerical
grid that has seven levels of mesh refinement, with a 2:1 refinement ratio,
centered on the BH. The finest level has a grid spacing of $dx\approx 0.02
M_0$. To estimate truncation error, we also perform the same calculation using
lower resolution. Details are given in the appendix. 

To characterize our solutions, we extract several relevant quantities.  Though
our solutions are neither stationary nor axisymmetric, it will be useful to
compute several mildly gauge dependent quantities with respect to the unit
normal to slices of constant coordinate time $n^a$, and the coordinate axial
vector $\hat{\phi}^a$ which would be the axisymmetric Killing vector in the
unperturbed BH spacetime.  From the stress-energy tensor of the massive vector
field $T^{ab}$, we can calculate the total energy $E$ and angular momentum $J$
from the volume integrals of $n_aT^a_t$ and $-n_a \hat{\phi}_b T^{ab}$,
respectively.  We also use the time component of the vector field
$\chi:=-n_aX^a$ as a convenient scalar quantity with respect to our spacetime
splitting.  During the evolution, we keep track of the BH apparent horizon, and
measure its area and its (approximate) angular momentum with respect to
$\hat{\phi}^a$. We combine these two quantities using the Christodoulou formula
to construct a measure of the total BH mass $M_{\rm BH}$.  To measure the GW
emission, we extract the Newman-Penrose scalar $\Psi_4$.

\begin{figure}
\begin{center}
\includegraphics[width=\columnwidth,draft=false]{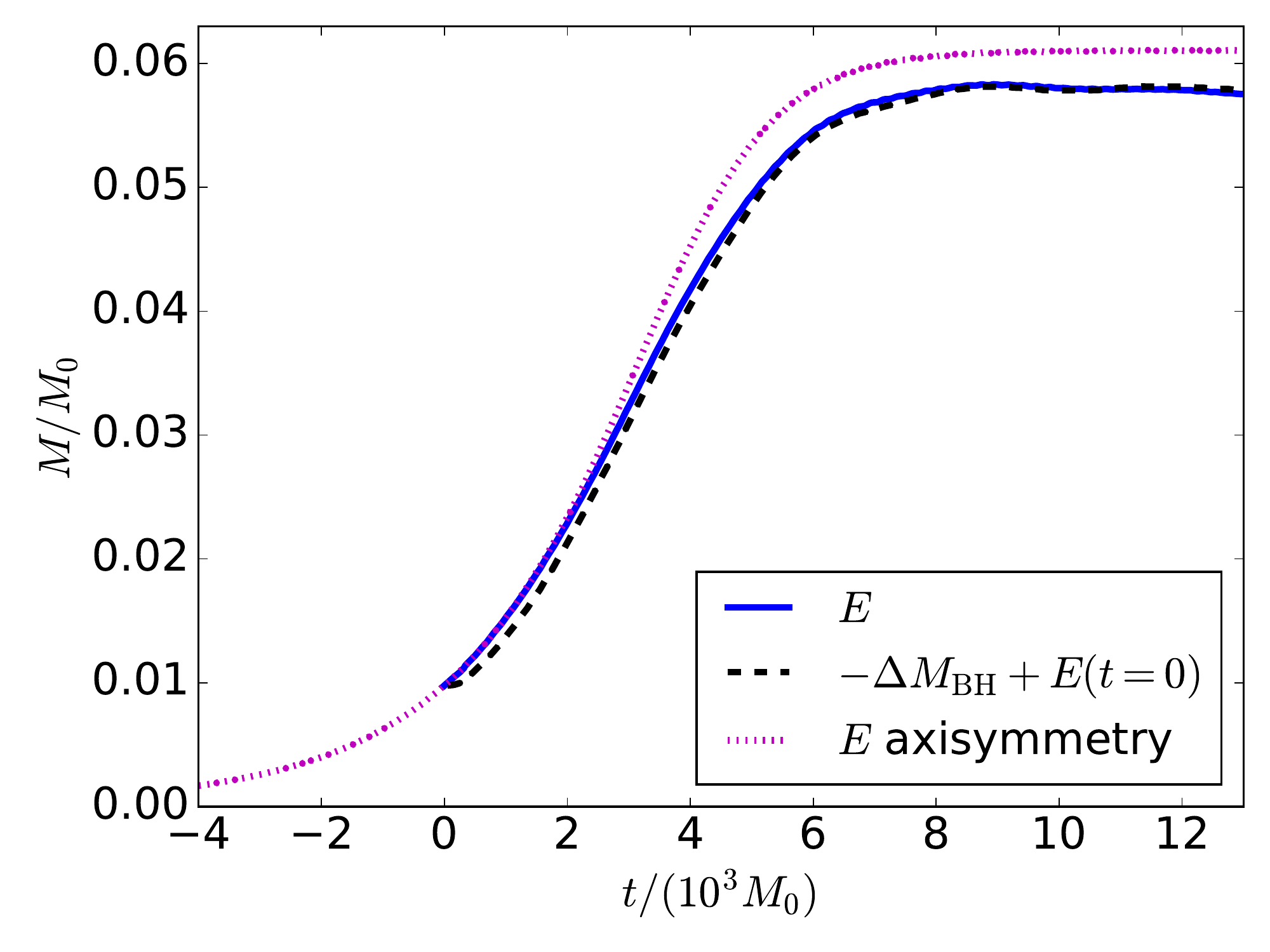}
\includegraphics[width=\columnwidth,draft=false]{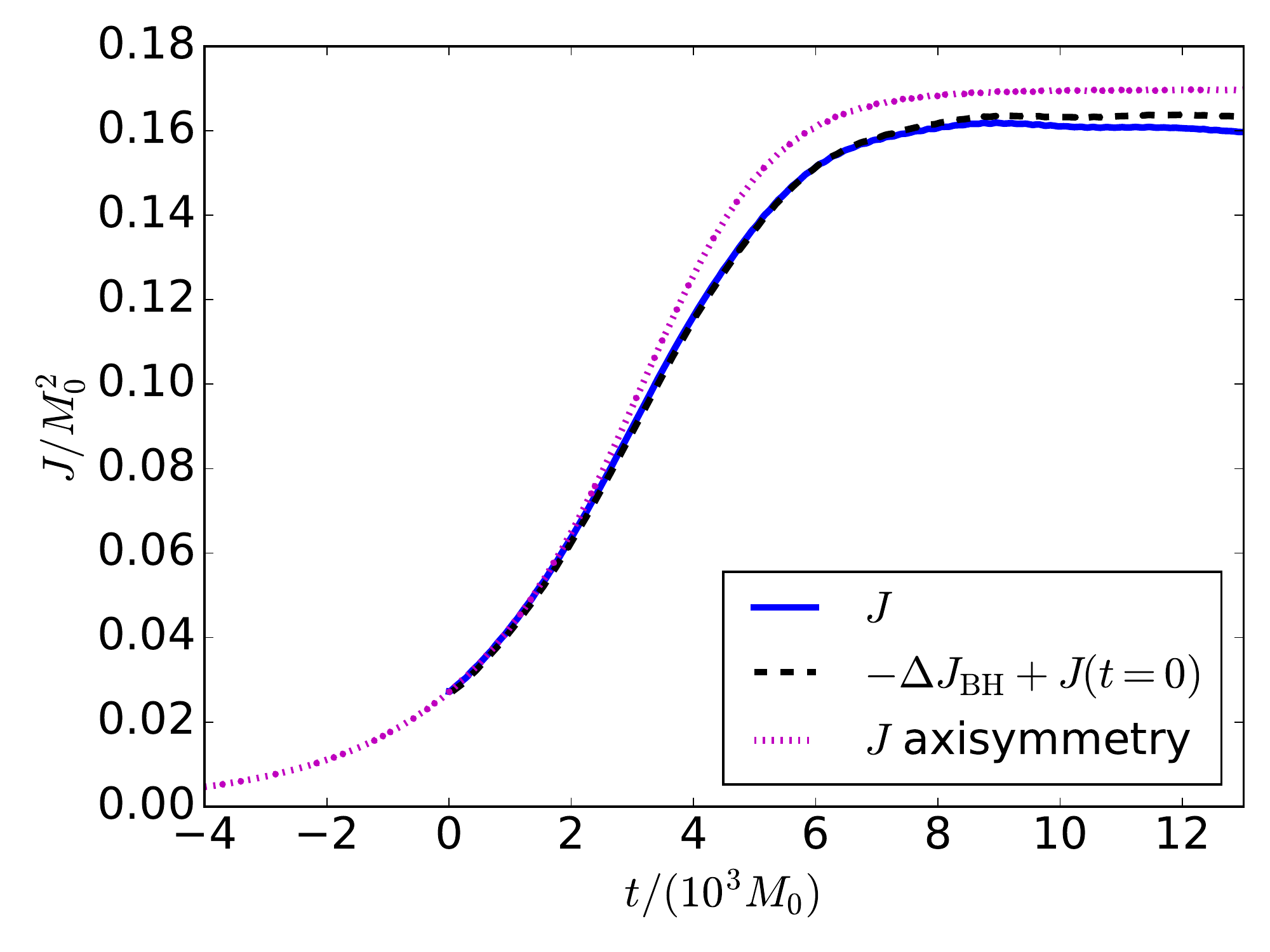}
\end{center}
\caption{
The energy (top) and angular momentum (bottom) in the Proca field
as a function of time (solid blue lines). In addition, we show the corresponding
decrease in the mass and angular momentum measured from the BH horizon, 
with initial offset as given by the initial Proca field values (dotted
black lines). For comparison, we also show the energy and angular momentum
in the Proca field from the axisymmetric evolutions described 
in Ref.~\cite{East:2017ovw}. The difference in the saturation values between
the two cases is predominantly due to the small change in the BH parameters
caused by solving the constraint equations when constructing the nonaxisymmetric initial data.
\label{fig:ej}
}
\end{figure}

\ssec{Results}%
We follow the boson cloud as it grows from $E/M_0\approx 0.01$
to a maximum value of $0.06$ over a timescale of $\sim 9\times 10^3M_0$,
and then subsequently decreases due to the emission of gravitational
radiation. This evolution of the cloud energy and angular momentum
closely matches the decrease in the BH mass and angular momentum, 
as shown in Fig.~\ref{fig:ej}. It is this spin-down of the 
BH that causes the instability to shut off---by 
$t=8\times 10^3M_0$, the BH horizon frequency is within $\sim2\%$
of the cloud oscillation frequency, and the condition for superradiance
is nearly saturated.
For comparison, we also show the 
evolution of the Proca field quantities from the corresponding
axisymmetric case. There is a small difference in the saturation
values of these quantities, primarily due to the shift
in BH parameters when constructing the initial data as described above.
Also, unlike the axisymmetric case, at late times $E$ and $J$ begin 
to decrease due to GW emission, as described below.   

\begin{figure}
\begin{center}
\includegraphics[width=\columnwidth,draft=false]{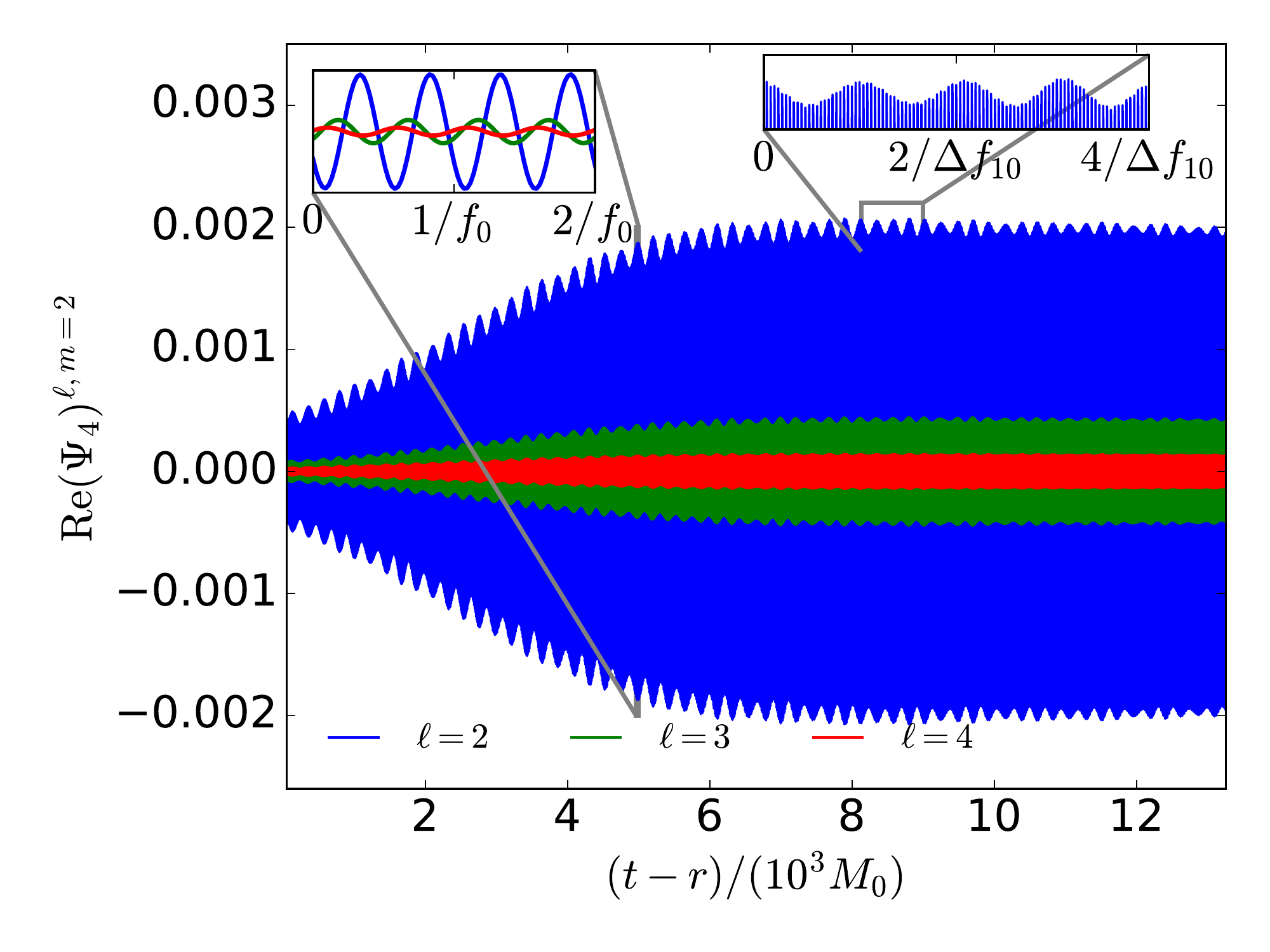}
\includegraphics[width=\columnwidth,draft=false]{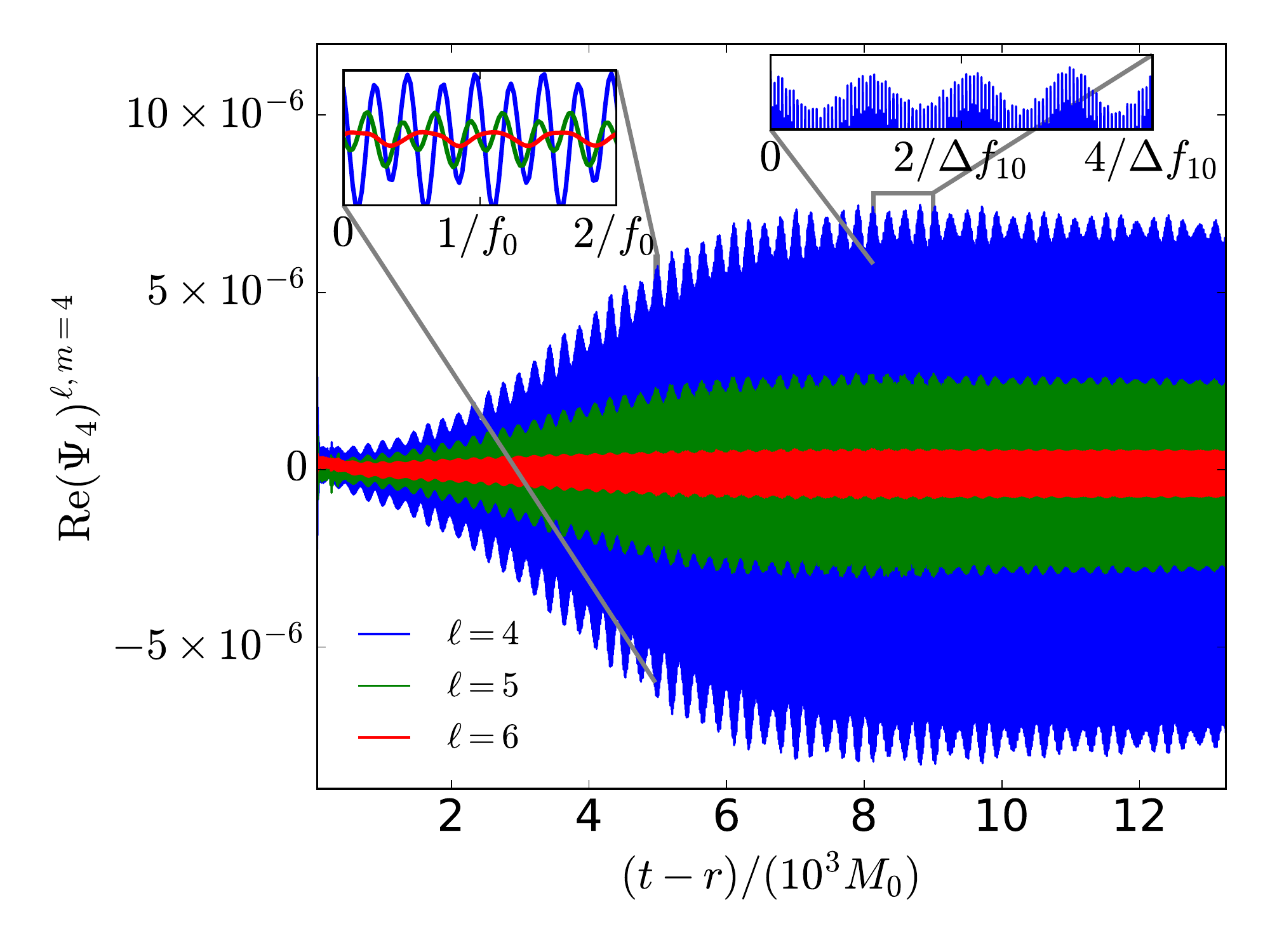}
\end{center}
\caption{
Spherical harmonics of the real part of $\Psi_4$, which encodes the GW signal.
The top panel shows the $2\leq \ell \leq 4$, $m=2$ components, while the the
bottom panel shows the $4\leq \ell \leq 6$, $m=4$ components.  The left-hand inset
demonstrates that the $m=2$ ($m=4$) signal is predominantly at 2 (4) times the
frequency of the most unstable mode $f_0=\omega_0/(2\pi)$, while the right-hand inset
demonstrates that there is a smaller variation in the amplitude due to 
beating of the first and second most unstable modes at frequency $\Delta f_{10}:=f_1-f_0$.
\label{fig:gw}
}
\end{figure}

The oscillations of the boson cloud produce GWs, as shown in Fig.~\ref{fig:gw}.
The GW signal is predominately quadrupolar, i.e.,  $\ell=|m|=2$ spin weight $-2$ spherical
harmonics, and has approximately twice the frequency of the dominant unstable
mode, $\omega_0\approx 0.359/M_0$ [the $(j,m,n)=(0,1,0)$ test field mode for
$\mu=0.4/M_0$ and $a=0.99$]~\cite{East:2017mrj}.  In the language of particle physics, this is
the ``annihilation" GW signal.  There is also a small modulation in the
amplitude that occurs with lower frequency which can be attributed to a beating
of the frequency of the dominant unstable mode ($n=0$) with a smaller amplitude
overtone\footnote{There are also 
other polarization modes in addition to the dominant $S=-1$. However, they grow
at a much slower rate than the first overtone~\cite{Dolan:2018dqv}.} 
($n=1$) with slightly higher frequency, $\Delta
\omega_{10}=\omega_1-\omega_0=0.029/M_0$~\cite{Dolan:2018dqv}, also
referred to as the ``transition" GW signal.  The amplitude of this secondary
mode is somewhat an artifact of our initial data, and the fact that we only
track the cloud through a small number of $e$-folds.  The growth rate of this
mode is approximately $1/4\times$ smaller than the dominant mode, and so would
be exponentially suppressed if the cloud grows for many e-folds (and will
also begin to decay at late times when $\omega_1$ is above the BH horizon frequency).  
However, such
a lower frequency transition GW signal might be relevant, in particular, if there
is part of the parameter space where an overtone mode has a faster growth rate
than the fundamental mode~\cite{Baryakhtar:2017ngi}, as can occur for the
scalar case~\cite{Yoshino:2015nsa}.

Figure~\ref{fig:gw} also shows that the GW signal has higher $\ell$ spherical
harmonic components, which is expected since the cloud is localized close to
the nonspherical, spinning BH. In the bottom panel we can see that there is also an $m=4$
component to the GWs at twice the frequency of the $m=2$ component.
The $m=4$ component has a smaller amplitude that grows quadratically with the 
cloud energy, consistent with being due to mode doubling effects. 
  
\begin{figure}
\begin{center}
\includegraphics[width=\columnwidth,draft=false]{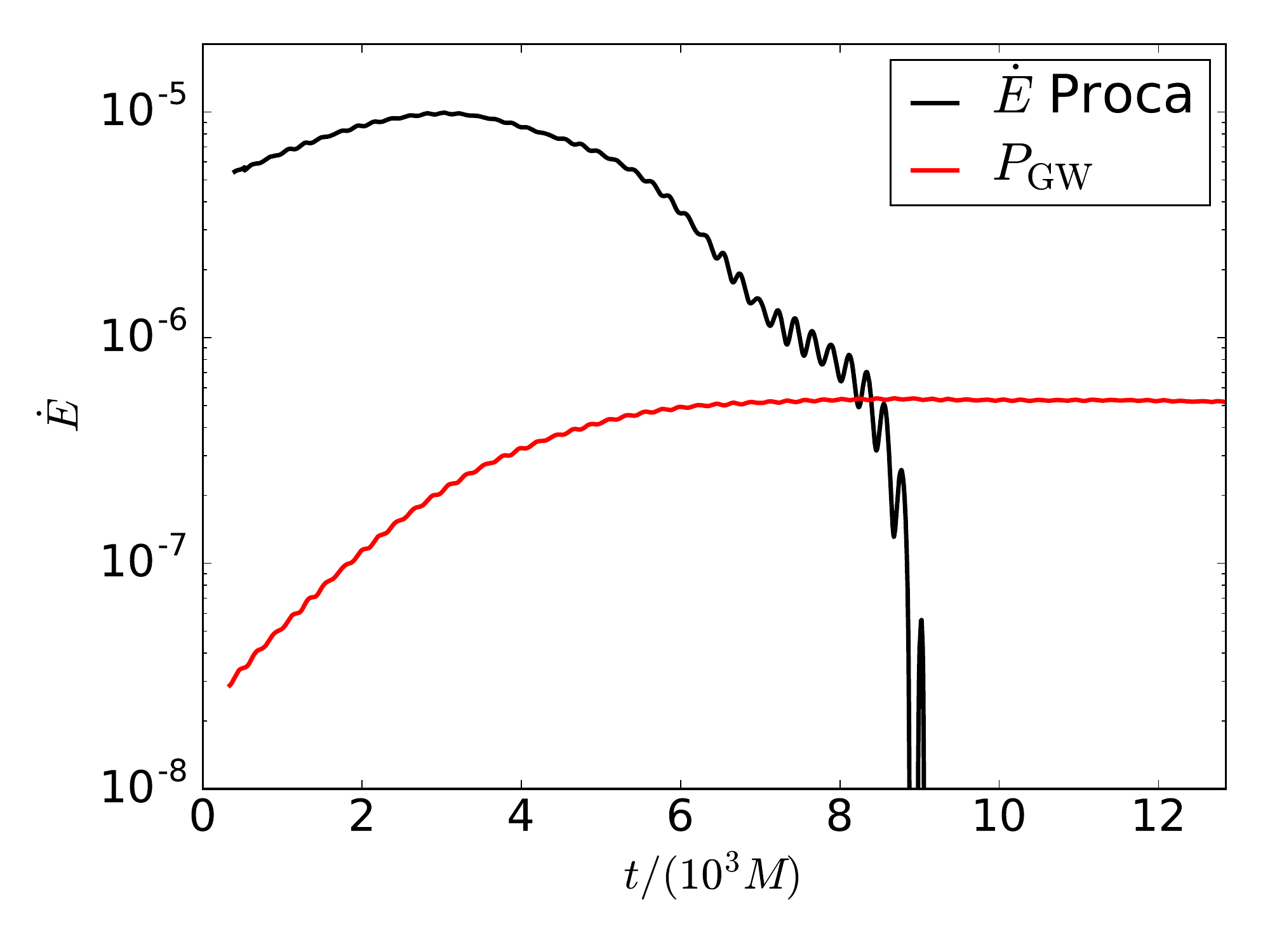}
\includegraphics[width=\columnwidth,draft=false]{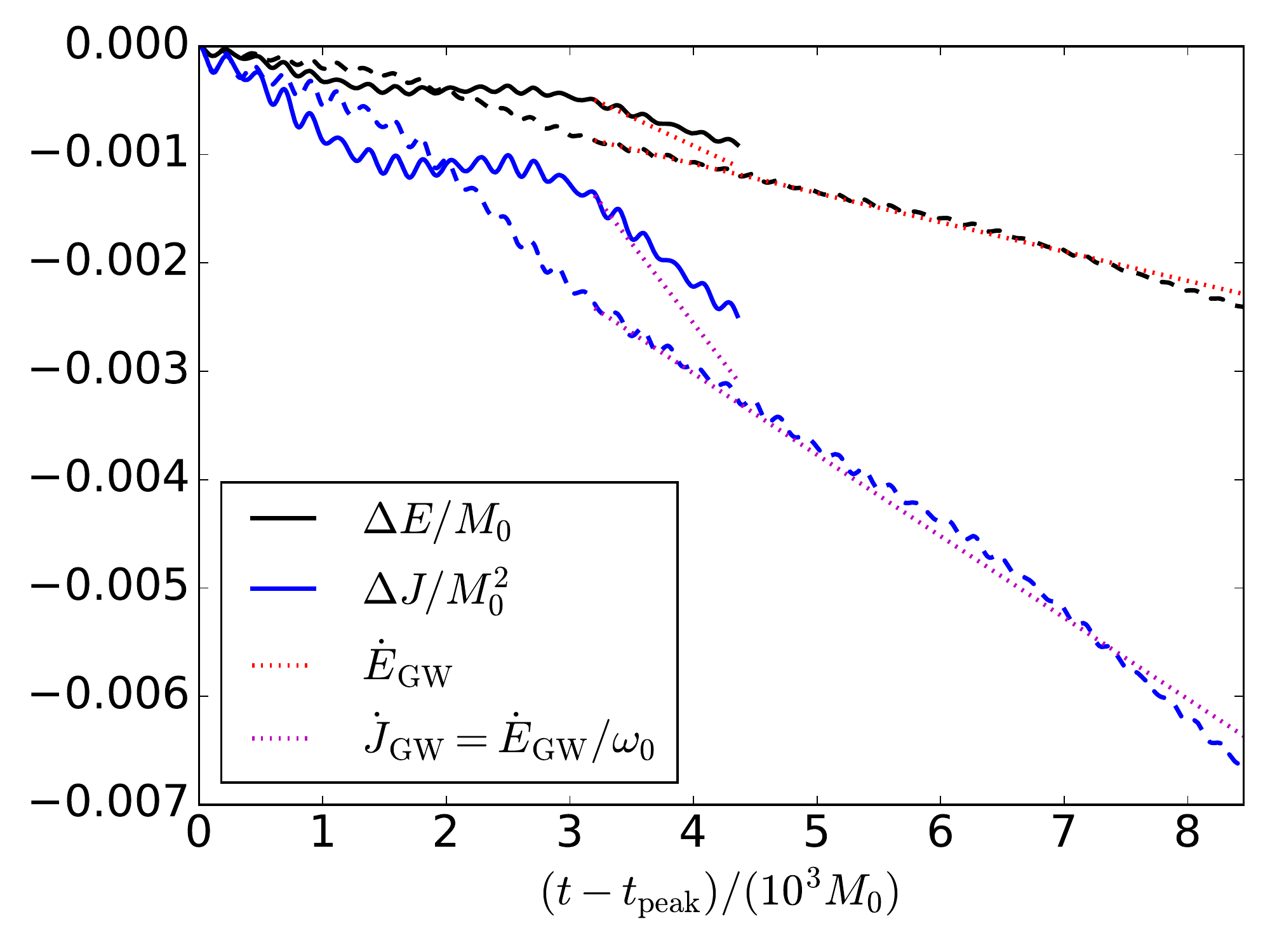}
\end{center}
\caption{
    Top: The time derivative of the (smoothed) Proca energy (black line) and the
GW power (red line). Bottom: The decrease in the Proca cloud energy
(black lines) and angular momentum (blue lines) from its peak value. The
dashed lines show the results from the lower resolution run, which is continued
for longer times. For comparison, we also show the rate of decrease expected
from GW dissipation (dotted red and magenta lines, respectively) for both resolutions.
\label{fig:power}
}
\end{figure}

The GW radiation is subdominant during the exponential growth phase of the
cloud. 
This is illustrated in the top panel of Fig.~\ref{fig:power}, where we show
the rate of increase of the boson cloud energy and the GW luminosity $P_{\rm
GW}$ as a function of time.
The value of $P_{\rm GW}/E^2$ found here when the cloud size is at maximum
mass matches the value found in the test field regime~\cite{East:2017mrj} to 
within the estimated numerical error (which is $\sim 17\%$). 
This peak value of the GW luminosity is $\sim20\times$ smaller than the
maximum growth rate of the Proca cloud.
However, eventually as the BH is spun down and the superradiant instability
saturates, the boson cloud begins to lose energy with a rate that roughly matches the GW
emission, as shown in the bottom panel. (At earlier times in the bottom panel of
Fig.\ref{fig:power}, the energy and angular momentum does briefly plateau, perhaps
due to the presence of the overtone mode.)
The cloud will dissipate on a characteristic timescale
of $t_{\rm GW}:=E/P_{\rm GW}\sim 10^{5}M_0$.

\begin{figure}
\begin{center}
\includegraphics[width=\columnwidth,draft=false]{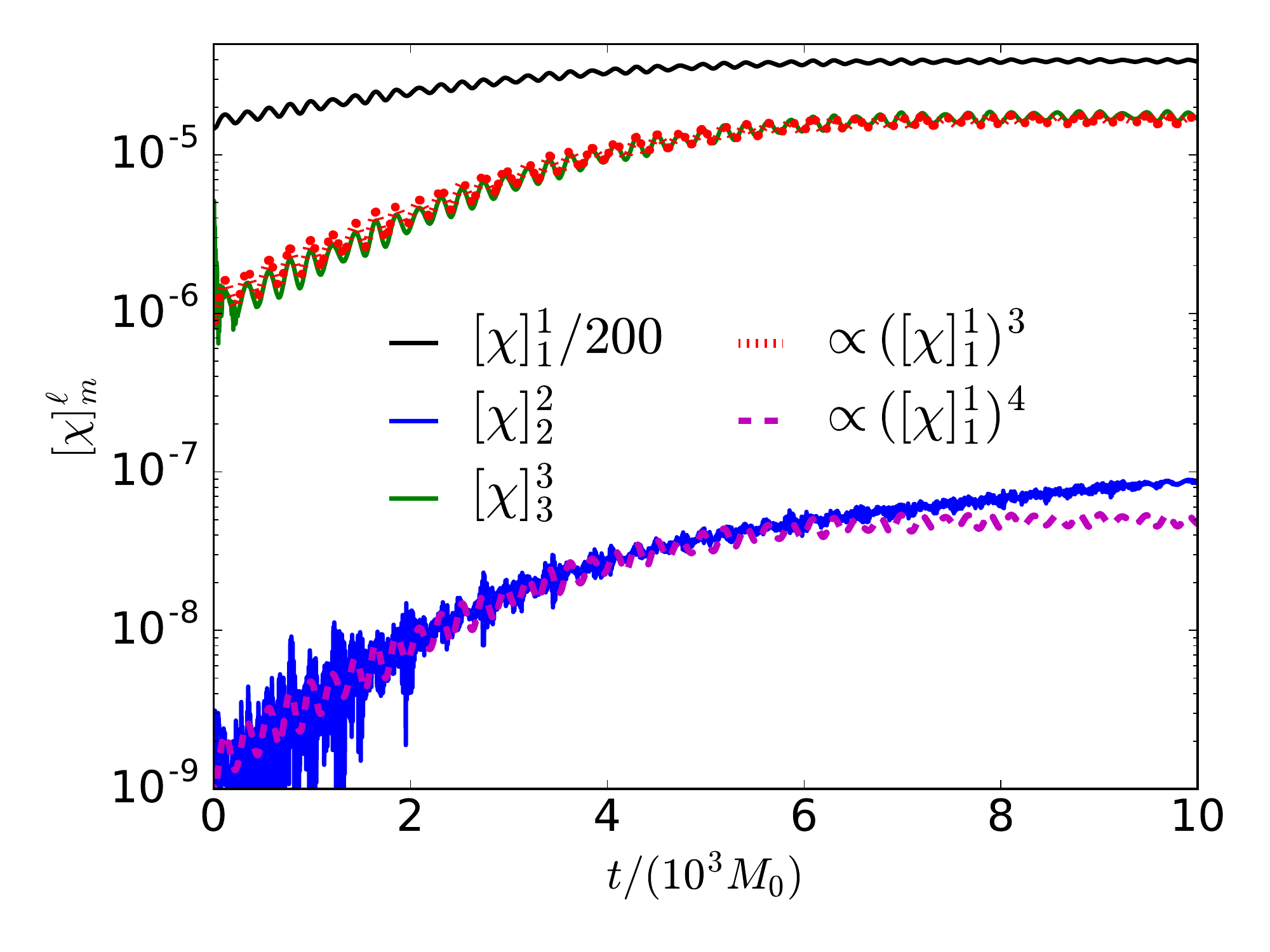}
\end{center}
\caption{
The $\ell=m=1$, 2, and 3 spherical harmonic components of $\chi$ at $r/M_0=6$
 versus time.  The dominant $\ell=m=1$ component has been scaled down
by a factor of 200 to fit on the plot.  We also show that the $\ell=m=3$ component 
appears proportional to the cube of the $\ell=m=1$ mode, and that the
$\ell=m=2$ component appears approximately proportional to the fourth power of the $\ell=m=1$
mode.
\label{fig:chi_phi}
}
\end{figure}

If we look at the vector field itself---in particular, $\chi$ as shown in
Fig.~\ref{fig:chi_phi}---in addition to the dominant $m=1$, there are higher $m$
components (predominantly $\ell=m$) at a much smaller amplitude (though as noted above, 
since there is no axisymmetry, such a decomposition is not free of gauge dependence).  
The increase
of these appear consistent with nonlinear mode coupling through gravity, with
the $m=3$ and $m=2$ mode increasing in proportion to the third and fourth power
of the $m=1$ mode, respectively.  We note that though there are superradiantly
unstable modes with higher $m$, which can continue to spin down the BH after
the saturation of the $m=1$ mode and subsequent dissipation of the cloud due to
GWs, their growth time is much longer than considered here ($>10^7M_0$).
However, nonlinear mode coupling could play a role in seeding such superradiant
modes at a higher amplitude.

\ssec{Conclusion}%
We have studied the development of the superradiant instability of a spinning
BH in the presence of a massive vector field, following the spin-down of the BH
and the growth of an oscillating boson cloud with significant mass and
attendant GW radiation.  We have found that the superradiant
instability can efficiently liberate a significant fraction of an isolated
spinning BH's mass, and convert it into GWs.  
As a consequence of the fact that, at saturation, the BH horizon has 
a frequency that matches the oscillating cloud, 
GW absorption or tidal heating effects are suppressed~\cite{Poisson:2004cw}.

Restoring astrophysical units suitable to a LIGO BH event to illustrate the
case considered here, the GW signal has a peak luminosity of $P_{\rm GW}\sim
2\times10^{53} {\rm \ erg/s}$, a characteristic timescale of $t_{\rm GW}\sim 30
{\rm \ sec} (M_0/60\ M_{\odot})$, and is predominantly at a frequency of $f_{\rm
GW}\approx (400{\rm \ Hz})(60\ M_{\odot}/M_0)$.  The amount of GW energy and
frequency is comparable to that of a BH merger---though the peak luminosity is
$\sim10^{-3}\times$ smaller---and such a signal would be visible by LIGO at
distances on the order of a Gpc.  This exact value of the vector boson mass is
in tension with reported BH spin measurements, though observations of higher BH
masses can probe new parts of the parameter space~\cite{Baryakhtar:2017ngi}.
We have focused on a maximally relativistic case, while for larger vector
bosonic Compton wavelengths (compared to the BH radius), as well as for the
case of a scalar boson, the growth rates and GW luminosity are smaller, and we
expect the adiabatic approximation to be even more accurate. These sources will
thus be promising targets for continuous GW
searches~\cite{Goncharov:2018ufi,Pierce:2018xmy}.  

Besides the spin-down of the BH leading to saturation of the instability, and
the dissipation of the boson cloud through GWs, we see subdominant nonlinear
effects, including evidence of mode coupling both in the vector field and in
the GWs, similar to that reported in Refs.~\cite{East:2013mfa,Chesler:2018txn}. It
would interesting for future work to investigate what role these could play in
speeding up the onset of higher azimuthal number superradiant instabilities,
which could liberate further rotational energy from the BH at late times.  The
large energy densities and high frequency oscillations of the boson clouds
exhibited here also suggest studying possible
observational signatures of coupling to an accretion disk or other matter.

\ssec{Acknowledgements}%
I thank Frans Pretorius, Helvi Witek, Huan Yang, and Miguel Zilhao for stimulating
discussions and comments on this work.  This research was supported in part by
Perimeter Institute for Theoretical Physics. Research at Perimeter Institute is
supported by the Government of Canada through the Department of Innovation,
Science and Economic Development Canada and by the Province of Ontario through
the Ministry of Research, Innovation and Science.  Computational resources were
provided by the Perseus cluster at Princeton University and by XSEDE under
Grant No. TG-PHY100053.

\bibliographystyle{apsrev4-1.bst}
\bibliography{ref}

\begin{thebibliography}{42}%
\makeatletter
\providecommand \@ifxundefined [1]{%
 \@ifx{#1\undefined}
}%
\providecommand \@ifnum [1]{%
 \ifnum #1\expandafter \@firstoftwo
 \else \expandafter \@secondoftwo
 \fi
}%
\providecommand \@ifx [1]{%
 \ifx #1\expandafter \@firstoftwo
 \else \expandafter \@secondoftwo
 \fi
}%
\providecommand \natexlab [1]{#1}%
\providecommand \enquote  [1]{``#1''}%
\providecommand \bibnamefont  [1]{#1}%
\providecommand \bibfnamefont [1]{#1}%
\providecommand \citenamefont [1]{#1}%
\providecommand \href@noop [0]{\@secondoftwo}%
\providecommand \href [0]{\begingroup \@sanitize@url \@href}%
\providecommand \@href[1]{\@@startlink{#1}\@@href}%
\providecommand \@@href[1]{\endgroup#1\@@endlink}%
\providecommand \@sanitize@url [0]{\catcode `\\12\catcode `\$12\catcode
  `\&12\catcode `\#12\catcode `\^12\catcode `\_12\catcode `\%12\relax}%
\providecommand \@@startlink[1]{}%
\providecommand \@@endlink[0]{}%
\providecommand \url  [0]{\begingroup\@sanitize@url \@url }%
\providecommand \@url [1]{\endgroup\@href {#1}{\urlprefix }}%
\providecommand \urlprefix  [0]{URL }%
\providecommand \Eprint [0]{\href }%
\providecommand \doibase [0]{http://dx.doi.org/}%
\providecommand \selectlanguage [0]{\@gobble}%
\providecommand \bibinfo  [0]{\@secondoftwo}%
\providecommand \bibfield  [0]{\@secondoftwo}%
\providecommand \translation [1]{[#1]}%
\providecommand \BibitemOpen [0]{}%
\providecommand \bibitemStop [0]{}%
\providecommand \bibitemNoStop [0]{.\EOS\space}%
\providecommand \EOS [0]{\spacefactor3000\relax}%
\providecommand \BibitemShut  [1]{\csname bibitem#1\endcsname}%
\let\auto@bib@innerbib\@empty
\bibitem [{\citenamefont {Abbott}\ \emph {et~al.}(2016)\citenamefont {Abbott}
  \emph {et~al.}}]{Abbott:2016blz}%
  \BibitemOpen
  \bibfield  {author} {\bibinfo {author} {\bibfnamefont {B.~P.}\ \bibnamefont
  {Abbott}} \emph {et~al.} (\bibinfo {collaboration} {Virgo, LIGO
  Scientific}),\ }\href {\doibase 10.1103/PhysRevLett.116.061102} {\bibfield
  {journal} {\bibinfo  {journal} {Phys. Rev. Lett.}\ }\textbf {\bibinfo
  {volume} {116}},\ \bibinfo {pages} {061102} (\bibinfo {year} {2016})},\
  \Eprint {http://arxiv.org/abs/1602.03837} {arXiv:1602.03837 [gr-qc]}
  \BibitemShut {NoStop}%
\bibitem [{\citenamefont {Weinberg}(1978)}]{Weinberg:1977ma}%
  \BibitemOpen
  \bibfield  {author} {\bibinfo {author} {\bibfnamefont {S.}~\bibnamefont
  {Weinberg}},\ }\href {\doibase 10.1103/PhysRevLett.40.223} {\bibfield
  {journal} {\bibinfo  {journal} {Phys.Rev.Lett.}\ }\textbf {\bibinfo {volume}
  {40}},\ \bibinfo {pages} {223} (\bibinfo {year} {1978})}\BibitemShut
  {NoStop}%
\bibitem [{\citenamefont {Arvanitaki}\ \emph {et~al.}(2010)\citenamefont
  {Arvanitaki}, \citenamefont {Dimopoulos}, \citenamefont {Dubovsky},
  \citenamefont {Kaloper},\ and\ \citenamefont
  {March-Russell}}]{Arvanitaki:2009fg}%
  \BibitemOpen
  \bibfield  {author} {\bibinfo {author} {\bibfnamefont {A.}~\bibnamefont
  {Arvanitaki}}, \bibinfo {author} {\bibfnamefont {S.}~\bibnamefont
  {Dimopoulos}}, \bibinfo {author} {\bibfnamefont {S.}~\bibnamefont
  {Dubovsky}}, \bibinfo {author} {\bibfnamefont {N.}~\bibnamefont {Kaloper}}, \
  and\ \bibinfo {author} {\bibfnamefont {J.}~\bibnamefont {March-Russell}},\
  }\href {\doibase 10.1103/PhysRevD.81.123530} {\bibfield  {journal} {\bibinfo
  {journal} {Phys. Rev.}\ }\textbf {\bibinfo {volume} {D81}},\ \bibinfo {pages}
  {123530} (\bibinfo {year} {2010})},\ \Eprint {http://arxiv.org/abs/0905.4720}
  {arXiv:0905.4720 [hep-th]} \BibitemShut {NoStop}%
\bibitem [{\citenamefont {Arvanitaki}\ and\ \citenamefont
  {Dubovsky}(2011)}]{Arvanitaki:2010sy}%
  \BibitemOpen
  \bibfield  {author} {\bibinfo {author} {\bibfnamefont {A.}~\bibnamefont
  {Arvanitaki}}\ and\ \bibinfo {author} {\bibfnamefont {S.}~\bibnamefont
  {Dubovsky}},\ }\href {\doibase 10.1103/PhysRevD.83.044026} {\bibfield
  {journal} {\bibinfo  {journal} {Phys. Rev.}\ }\textbf {\bibinfo {volume}
  {D83}},\ \bibinfo {pages} {044026} (\bibinfo {year} {2011})},\ \Eprint
  {http://arxiv.org/abs/1004.3558} {arXiv:1004.3558 [hep-th]} \BibitemShut
  {NoStop}%
\bibitem [{\citenamefont {Holdom}(1986)}]{Holdom:1985ag}%
  \BibitemOpen
  \bibfield  {author} {\bibinfo {author} {\bibfnamefont {B.}~\bibnamefont
  {Holdom}},\ }\href {\doibase 10.1016/0370-2693(86)91377-8} {\bibfield
  {journal} {\bibinfo  {journal} {Phys. Lett.}\ }\textbf {\bibinfo {volume}
  {B166}},\ \bibinfo {pages} {196} (\bibinfo {year} {1986})}\BibitemShut
  {NoStop}%
\bibitem [{\citenamefont {Cicoli}\ \emph {et~al.}(2011)\citenamefont {Cicoli},
  \citenamefont {Goodsell}, \citenamefont {Jaeckel},\ and\ \citenamefont
  {Ringwald}}]{Cicoli:2011yh}%
  \BibitemOpen
  \bibfield  {author} {\bibinfo {author} {\bibfnamefont {M.}~\bibnamefont
  {Cicoli}}, \bibinfo {author} {\bibfnamefont {M.}~\bibnamefont {Goodsell}},
  \bibinfo {author} {\bibfnamefont {J.}~\bibnamefont {Jaeckel}}, \ and\
  \bibinfo {author} {\bibfnamefont {A.}~\bibnamefont {Ringwald}},\ }\href
  {\doibase 10.1007/JHEP07(2011)114} {\bibfield  {journal} {\bibinfo  {journal}
  {JHEP}\ }\textbf {\bibinfo {volume} {07}},\ \bibinfo {pages} {114} (\bibinfo
  {year} {2011})},\ \Eprint {http://arxiv.org/abs/1103.3705} {arXiv:1103.3705
  [hep-th]} \BibitemShut {NoStop}%
\bibitem [{\citenamefont {Damour}\ \emph {et~al.}(1976)\citenamefont {Damour},
  \citenamefont {Deruelle},\ and\ \citenamefont {Ruffini}}]{Damour:1976}%
  \BibitemOpen
  \bibfield  {author} {\bibinfo {author} {\bibfnamefont {T.}~\bibnamefont
  {Damour}}, \bibinfo {author} {\bibfnamefont {N.}~\bibnamefont {Deruelle}}, \
  and\ \bibinfo {author} {\bibfnamefont {R.}~\bibnamefont {Ruffini}},\ }\href
  {\doibase 10.1007/BF02725534} {\bibfield  {journal} {\bibinfo  {journal}
  {Lettere Al Nuovo Cimento Series 2}\ }\textbf {\bibinfo {volume} {15}},\
  \bibinfo {pages} {257} (\bibinfo {year} {1976})}\BibitemShut {NoStop}%
\bibitem [{\citenamefont {Detweiler}(1980)}]{Detweiler:1980uk}%
  \BibitemOpen
  \bibfield  {author} {\bibinfo {author} {\bibfnamefont {S.~L.}\ \bibnamefont
  {Detweiler}},\ }\href {\doibase 10.1103/PhysRevD.22.2323} {\bibfield
  {journal} {\bibinfo  {journal} {Phys.Rev.}\ }\textbf {\bibinfo {volume}
  {D22}},\ \bibinfo {pages} {2323} (\bibinfo {year} {1980})}\BibitemShut
  {NoStop}%
\bibitem [{\citenamefont {Zouros}\ and\ \citenamefont
  {Eardley}(1979)}]{Zouros:1979iw}%
  \BibitemOpen
  \bibfield  {author} {\bibinfo {author} {\bibfnamefont {T.}~\bibnamefont
  {Zouros}}\ and\ \bibinfo {author} {\bibfnamefont {D.}~\bibnamefont
  {Eardley}},\ }\href {\doibase 10.1016/0003-4916(79)90237-9} {\bibfield
  {journal} {\bibinfo  {journal} {Annals Phys.}\ }\textbf {\bibinfo {volume}
  {118}},\ \bibinfo {pages} {139} (\bibinfo {year} {1979})}\BibitemShut
  {NoStop}%
\bibitem [{\citenamefont {Dolan}(2013)}]{Dolan:2012yt}%
  \BibitemOpen
  \bibfield  {author} {\bibinfo {author} {\bibfnamefont {S.~R.}\ \bibnamefont
  {Dolan}},\ }\href {\doibase 10.1103/PhysRevD.87.124026} {\bibfield  {journal}
  {\bibinfo  {journal} {Phys. Rev.}\ }\textbf {\bibinfo {volume} {D87}},\
  \bibinfo {pages} {124026} (\bibinfo {year} {2013})},\ \Eprint
  {http://arxiv.org/abs/1212.1477} {arXiv:1212.1477 [gr-qc]} \BibitemShut
  {NoStop}%
\bibitem [{\citenamefont {{Brito}}\ \emph {et~al.}(2015)\citenamefont
  {{Brito}}, \citenamefont {{Cardoso}},\ and\ \citenamefont
  {{Pani}}}]{2015CQGra..32m4001B}%
  \BibitemOpen
  \bibfield  {author} {\bibinfo {author} {\bibfnamefont {R.}~\bibnamefont
  {{Brito}}}, \bibinfo {author} {\bibfnamefont {V.}~\bibnamefont {{Cardoso}}},
  \ and\ \bibinfo {author} {\bibfnamefont {P.}~\bibnamefont {{Pani}}},\ }\href
  {\doibase 10.1088/0264-9381/32/13/134001} {\bibfield  {journal} {\bibinfo
  {journal} {Classical and Quantum Gravity}\ }\textbf {\bibinfo {volume}
  {32}},\ \bibinfo {eid} {134001} (\bibinfo {year} {2015})},\ \Eprint
  {http://arxiv.org/abs/1411.0686} {arXiv:1411.0686 [gr-qc]} \BibitemShut
  {NoStop}%
\bibitem [{\citenamefont {East}\ and\ \citenamefont
  {Pretorius}(2017)}]{East:2017ovw}%
  \BibitemOpen
  \bibfield  {author} {\bibinfo {author} {\bibfnamefont {W.~E.}\ \bibnamefont
  {East}}\ and\ \bibinfo {author} {\bibfnamefont {F.}~\bibnamefont
  {Pretorius}},\ }\href {\doibase 10.1103/PhysRevLett.119.041101} {\bibfield
  {journal} {\bibinfo  {journal} {Phys. Rev. Lett.}\ }\textbf {\bibinfo
  {volume} {119}},\ \bibinfo {pages} {041101} (\bibinfo {year} {2017})},\
  \Eprint {http://arxiv.org/abs/1704.04791} {arXiv:1704.04791 [gr-qc]}
  \BibitemShut {NoStop}%
\bibitem [{\citenamefont {Arvanitaki}\ \emph {et~al.}(2015)\citenamefont
  {Arvanitaki}, \citenamefont {Baryakhtar},\ and\ \citenamefont
  {Huang}}]{Arvanitaki:2014wva}%
  \BibitemOpen
  \bibfield  {author} {\bibinfo {author} {\bibfnamefont {A.}~\bibnamefont
  {Arvanitaki}}, \bibinfo {author} {\bibfnamefont {M.}~\bibnamefont
  {Baryakhtar}}, \ and\ \bibinfo {author} {\bibfnamefont {X.}~\bibnamefont
  {Huang}},\ }\href {\doibase 10.1103/PhysRevD.91.084011} {\bibfield  {journal}
  {\bibinfo  {journal} {Phys. Rev.}\ }\textbf {\bibinfo {volume} {D91}},\
  \bibinfo {pages} {084011} (\bibinfo {year} {2015})},\ \Eprint
  {http://arxiv.org/abs/1411.2263} {arXiv:1411.2263 [hep-ph]} \BibitemShut
  {NoStop}%
\bibitem [{\citenamefont {Arvanitaki}\ \emph {et~al.}(2017)\citenamefont
  {Arvanitaki}, \citenamefont {Baryakhtar}, \citenamefont {Dimopoulos},
  \citenamefont {Dubovsky},\ and\ \citenamefont
  {Lasenby}}]{Arvanitaki:2016qwi}%
  \BibitemOpen
  \bibfield  {author} {\bibinfo {author} {\bibfnamefont {A.}~\bibnamefont
  {Arvanitaki}}, \bibinfo {author} {\bibfnamefont {M.}~\bibnamefont
  {Baryakhtar}}, \bibinfo {author} {\bibfnamefont {S.}~\bibnamefont
  {Dimopoulos}}, \bibinfo {author} {\bibfnamefont {S.}~\bibnamefont
  {Dubovsky}}, \ and\ \bibinfo {author} {\bibfnamefont {R.}~\bibnamefont
  {Lasenby}},\ }\href {\doibase 10.1103/PhysRevD.95.043001} {\bibfield
  {journal} {\bibinfo  {journal} {Phys. Rev.}\ }\textbf {\bibinfo {volume}
  {D95}},\ \bibinfo {pages} {043001} (\bibinfo {year} {2017})},\ \Eprint
  {http://arxiv.org/abs/1604.03958} {arXiv:1604.03958 [hep-ph]} \BibitemShut
  {NoStop}%
\bibitem [{\citenamefont {Baryakhtar}\ \emph {et~al.}(2017)\citenamefont
  {Baryakhtar}, \citenamefont {Lasenby},\ and\ \citenamefont
  {Teo}}]{Baryakhtar:2017ngi}%
  \BibitemOpen
  \bibfield  {author} {\bibinfo {author} {\bibfnamefont {M.}~\bibnamefont
  {Baryakhtar}}, \bibinfo {author} {\bibfnamefont {R.}~\bibnamefont {Lasenby}},
  \ and\ \bibinfo {author} {\bibfnamefont {M.}~\bibnamefont {Teo}},\ }\href
  {\doibase 10.1103/PhysRevD.96.035019} {\bibfield  {journal} {\bibinfo
  {journal} {Phys. Rev.}\ }\textbf {\bibinfo {volume} {D96}},\ \bibinfo {pages}
  {035019} (\bibinfo {year} {2017})},\ \Eprint
  {http://arxiv.org/abs/1704.05081} {arXiv:1704.05081 [hep-ph]} \BibitemShut
  {NoStop}%
\bibitem [{\citenamefont {Baumann}\ \emph {et~al.}(2018)\citenamefont
  {Baumann}, \citenamefont {Chia},\ and\ \citenamefont
  {Porto}}]{Baumann:2018vus}%
  \BibitemOpen
  \bibfield  {author} {\bibinfo {author} {\bibfnamefont {D.}~\bibnamefont
  {Baumann}}, \bibinfo {author} {\bibfnamefont {H.~S.}\ \bibnamefont {Chia}}, \
  and\ \bibinfo {author} {\bibfnamefont {R.~A.}\ \bibnamefont {Porto}},\
  }\href@noop {} {\  (\bibinfo {year} {2018})},\ \Eprint
  {http://arxiv.org/abs/1804.03208} {arXiv:1804.03208 [gr-qc]} \BibitemShut
  {NoStop}%
\bibitem [{\citenamefont {Brito}\ \emph
  {et~al.}(2017{\natexlab{a}})\citenamefont {Brito}, \citenamefont {Ghosh},
  \citenamefont {Barausse}, \citenamefont {Berti}, \citenamefont {Cardoso},
  \citenamefont {Dvorkin}, \citenamefont {Klein},\ and\ \citenamefont
  {Pani}}]{Brito:2017zvb}%
  \BibitemOpen
  \bibfield  {author} {\bibinfo {author} {\bibfnamefont {R.}~\bibnamefont
  {Brito}}, \bibinfo {author} {\bibfnamefont {S.}~\bibnamefont {Ghosh}},
  \bibinfo {author} {\bibfnamefont {E.}~\bibnamefont {Barausse}}, \bibinfo
  {author} {\bibfnamefont {E.}~\bibnamefont {Berti}}, \bibinfo {author}
  {\bibfnamefont {V.}~\bibnamefont {Cardoso}}, \bibinfo {author} {\bibfnamefont
  {I.}~\bibnamefont {Dvorkin}}, \bibinfo {author} {\bibfnamefont
  {A.}~\bibnamefont {Klein}}, \ and\ \bibinfo {author} {\bibfnamefont
  {P.}~\bibnamefont {Pani}},\ }\href {\doibase 10.1103/PhysRevD.96.064050}
  {\bibfield  {journal} {\bibinfo  {journal} {Phys. Rev.}\ }\textbf {\bibinfo
  {volume} {D96}},\ \bibinfo {pages} {064050} (\bibinfo {year}
  {2017}{\natexlab{a}})},\ \Eprint {http://arxiv.org/abs/1706.06311}
  {arXiv:1706.06311 [gr-qc]} \BibitemShut {NoStop}%
\bibitem [{\citenamefont {Brito}\ \emph
  {et~al.}(2017{\natexlab{b}})\citenamefont {Brito}, \citenamefont {Ghosh},
  \citenamefont {Barausse}, \citenamefont {Berti}, \citenamefont {Cardoso},
  \citenamefont {Dvorkin}, \citenamefont {Klein},\ and\ \citenamefont
  {Pani}}]{Brito:2017wnc}%
  \BibitemOpen
  \bibfield  {author} {\bibinfo {author} {\bibfnamefont {R.}~\bibnamefont
  {Brito}}, \bibinfo {author} {\bibfnamefont {S.}~\bibnamefont {Ghosh}},
  \bibinfo {author} {\bibfnamefont {E.}~\bibnamefont {Barausse}}, \bibinfo
  {author} {\bibfnamefont {E.}~\bibnamefont {Berti}}, \bibinfo {author}
  {\bibfnamefont {V.}~\bibnamefont {Cardoso}}, \bibinfo {author} {\bibfnamefont
  {I.}~\bibnamefont {Dvorkin}}, \bibinfo {author} {\bibfnamefont
  {A.}~\bibnamefont {Klein}}, \ and\ \bibinfo {author} {\bibfnamefont
  {P.}~\bibnamefont {Pani}},\ }\href {\doibase 10.1103/PhysRevLett.119.131101}
  {\bibfield  {journal} {\bibinfo  {journal} {Phys. Rev. Lett.}\ }\textbf
  {\bibinfo {volume} {119}},\ \bibinfo {pages} {131101} (\bibinfo {year}
  {2017}{\natexlab{b}})},\ \Eprint {http://arxiv.org/abs/1706.05097}
  {arXiv:1706.05097 [gr-qc]} \BibitemShut {NoStop}%
\bibitem [{\citenamefont {East}(2017)}]{East:2017mrj}%
  \BibitemOpen
  \bibfield  {author} {\bibinfo {author} {\bibfnamefont {W.~E.}\ \bibnamefont
  {East}},\ }\href {\doibase 10.1103/PhysRevD.96.024004} {\bibfield  {journal}
  {\bibinfo  {journal} {Phys. Rev.}\ }\textbf {\bibinfo {volume} {D96}},\
  \bibinfo {pages} {024004} (\bibinfo {year} {2017})},\ \Eprint
  {http://arxiv.org/abs/1705.01544} {arXiv:1705.01544 [gr-qc]} \BibitemShut
  {NoStop}%
\bibitem [{\citenamefont {Pani}\ \emph
  {et~al.}(2012{\natexlab{a}})\citenamefont {Pani}, \citenamefont {Cardoso},
  \citenamefont {Gualtieri}, \citenamefont {Berti},\ and\ \citenamefont
  {Ishibashi}}]{Pani:2012bp}%
  \BibitemOpen
  \bibfield  {author} {\bibinfo {author} {\bibfnamefont {P.}~\bibnamefont
  {Pani}}, \bibinfo {author} {\bibfnamefont {V.}~\bibnamefont {Cardoso}},
  \bibinfo {author} {\bibfnamefont {L.}~\bibnamefont {Gualtieri}}, \bibinfo
  {author} {\bibfnamefont {E.}~\bibnamefont {Berti}}, \ and\ \bibinfo {author}
  {\bibfnamefont {A.}~\bibnamefont {Ishibashi}},\ }\href {\doibase
  10.1103/PhysRevD.86.104017} {\bibfield  {journal} {\bibinfo  {journal} {Phys.
  Rev.}\ }\textbf {\bibinfo {volume} {D86}},\ \bibinfo {pages} {104017}
  (\bibinfo {year} {2012}{\natexlab{a}})},\ \Eprint
  {http://arxiv.org/abs/1209.0773} {arXiv:1209.0773 [gr-qc]} \BibitemShut
  {NoStop}%
\bibitem [{\citenamefont {Pani}\ \emph
  {et~al.}(2012{\natexlab{b}})\citenamefont {Pani}, \citenamefont {Cardoso},
  \citenamefont {Gualtieri}, \citenamefont {Berti},\ and\ \citenamefont
  {Ishibashi}}]{Pani:2012vp}%
  \BibitemOpen
  \bibfield  {author} {\bibinfo {author} {\bibfnamefont {P.}~\bibnamefont
  {Pani}}, \bibinfo {author} {\bibfnamefont {V.}~\bibnamefont {Cardoso}},
  \bibinfo {author} {\bibfnamefont {L.}~\bibnamefont {Gualtieri}}, \bibinfo
  {author} {\bibfnamefont {E.}~\bibnamefont {Berti}}, \ and\ \bibinfo {author}
  {\bibfnamefont {A.}~\bibnamefont {Ishibashi}},\ }\href {\doibase
  10.1103/PhysRevLett.109.131102} {\bibfield  {journal} {\bibinfo  {journal}
  {Phys. Rev. Lett.}\ }\textbf {\bibinfo {volume} {109}},\ \bibinfo {pages}
  {131102} (\bibinfo {year} {2012}{\natexlab{b}})},\ \Eprint
  {http://arxiv.org/abs/1209.0465} {arXiv:1209.0465 [gr-qc]} \BibitemShut
  {NoStop}%
\bibitem [{\citenamefont {Endlich}\ and\ \citenamefont
  {Penco}(2017)}]{Endlich:2016jgc}%
  \BibitemOpen
  \bibfield  {author} {\bibinfo {author} {\bibfnamefont {S.}~\bibnamefont
  {Endlich}}\ and\ \bibinfo {author} {\bibfnamefont {R.}~\bibnamefont
  {Penco}},\ }\href {\doibase 10.1007/JHEP05(2017)052} {\bibfield  {journal}
  {\bibinfo  {journal} {JHEP}\ }\textbf {\bibinfo {volume} {05}},\ \bibinfo
  {pages} {052} (\bibinfo {year} {2017})},\ \Eprint
  {http://arxiv.org/abs/1609.06723} {arXiv:1609.06723 [hep-th]} \BibitemShut
  {NoStop}%
\bibitem [{\citenamefont {Witek}\ \emph {et~al.}(2013)\citenamefont {Witek},
  \citenamefont {Cardoso}, \citenamefont {Ishibashi},\ and\ \citenamefont
  {Sperhake}}]{Witek:2012tr}%
  \BibitemOpen
  \bibfield  {author} {\bibinfo {author} {\bibfnamefont {H.}~\bibnamefont
  {Witek}}, \bibinfo {author} {\bibfnamefont {V.}~\bibnamefont {Cardoso}},
  \bibinfo {author} {\bibfnamefont {A.}~\bibnamefont {Ishibashi}}, \ and\
  \bibinfo {author} {\bibfnamefont {U.}~\bibnamefont {Sperhake}},\ }\href
  {\doibase 10.1103/PhysRevD.87.043513} {\bibfield  {journal} {\bibinfo
  {journal} {Phys. Rev.}\ }\textbf {\bibinfo {volume} {D87}},\ \bibinfo {pages}
  {043513} (\bibinfo {year} {2013})},\ \Eprint {http://arxiv.org/abs/1212.0551}
  {arXiv:1212.0551 [gr-qc]} \BibitemShut {NoStop}%
\bibitem [{\citenamefont {Cardoso}\ \emph {et~al.}(2018)\citenamefont
  {Cardoso}, \citenamefont {Dias}, \citenamefont {Hartnett}, \citenamefont
  {Middleton}, \citenamefont {Pani},\ and\ \citenamefont
  {Santos}}]{Cardoso:2018tly}%
  \BibitemOpen
  \bibfield  {author} {\bibinfo {author} {\bibfnamefont {V.}~\bibnamefont
  {Cardoso}}, \bibinfo {author} {\bibfnamefont {O.~J.~C.}\ \bibnamefont
  {Dias}}, \bibinfo {author} {\bibfnamefont {G.~S.}\ \bibnamefont {Hartnett}},
  \bibinfo {author} {\bibfnamefont {M.}~\bibnamefont {Middleton}}, \bibinfo
  {author} {\bibfnamefont {P.}~\bibnamefont {Pani}}, \ and\ \bibinfo {author}
  {\bibfnamefont {J.~E.}\ \bibnamefont {Santos}},\ }\href {\doibase
  10.1088/1475-7516/2018/03/043} {\bibfield  {journal} {\bibinfo  {journal}
  {JCAP}\ }\textbf {\bibinfo {volume} {1803}},\ \bibinfo {pages} {043}
  (\bibinfo {year} {2018})},\ \Eprint {http://arxiv.org/abs/1801.01420}
  {arXiv:1801.01420 [gr-qc]} \BibitemShut {NoStop}%
\bibitem [{\citenamefont {Frolov}\ \emph {et~al.}(2018)\citenamefont {Frolov},
  \citenamefont {Krtous}, \citenamefont {Kubiznak},\ and\ \citenamefont
  {Santos}}]{Frolov:2018ezx}%
  \BibitemOpen
  \bibfield  {author} {\bibinfo {author} {\bibfnamefont {V.~P.}\ \bibnamefont
  {Frolov}}, \bibinfo {author} {\bibfnamefont {P.}~\bibnamefont {Krtous}},
  \bibinfo {author} {\bibfnamefont {D.}~\bibnamefont {Kubiznak}}, \ and\
  \bibinfo {author} {\bibfnamefont {J.~E.}\ \bibnamefont {Santos}},\
  }\href@noop {} {\  (\bibinfo {year} {2018})},\ \Eprint
  {http://arxiv.org/abs/1804.00030} {arXiv:1804.00030 [hep-th]} \BibitemShut
  {NoStop}%
\bibitem [{\citenamefont {Dolan}(2018)}]{Dolan:2018dqv}%
  \BibitemOpen
  \bibfield  {author} {\bibinfo {author} {\bibfnamefont {S.~R.}\ \bibnamefont
  {Dolan}},\ }\href@noop {} {\  (\bibinfo {year} {2018})},\ \Eprint
  {http://arxiv.org/abs/1806.01604} {arXiv:1806.01604 [gr-qc]} \BibitemShut
  {NoStop}%
\bibitem [{\citenamefont {East}\ \emph {et~al.}(2014)\citenamefont {East},
  \citenamefont {Ramazanoğlu},\ and\ \citenamefont
  {Pretorius}}]{East:2013mfa}%
  \BibitemOpen
  \bibfield  {author} {\bibinfo {author} {\bibfnamefont {W.~E.}\ \bibnamefont
  {East}}, \bibinfo {author} {\bibfnamefont {F.~M.}\ \bibnamefont
  {Ramazanoğlu}}, \ and\ \bibinfo {author} {\bibfnamefont {F.}~\bibnamefont
  {Pretorius}},\ }\href {\doibase 10.1103/PhysRevD.89.061503} {\bibfield
  {journal} {\bibinfo  {journal} {Phys. Rev.}\ }\textbf {\bibinfo {volume}
  {D89}},\ \bibinfo {pages} {061503} (\bibinfo {year} {2014})},\ \Eprint
  {http://arxiv.org/abs/1312.4529} {arXiv:1312.4529 [gr-qc]} \BibitemShut
  {NoStop}%
\bibitem [{\citenamefont {Chesler}\ and\ \citenamefont
  {Lowe}(2018)}]{Chesler:2018txn}%
  \BibitemOpen
  \bibfield  {author} {\bibinfo {author} {\bibfnamefont {P.~M.}\ \bibnamefont
  {Chesler}}\ and\ \bibinfo {author} {\bibfnamefont {D.~A.}\ \bibnamefont
  {Lowe}},\ }\href@noop {} {\  (\bibinfo {year} {2018})},\ \Eprint
  {http://arxiv.org/abs/1801.09711} {arXiv:1801.09711 [gr-qc]} \BibitemShut
  {NoStop}%
\bibitem [{\citenamefont {Sanchis-Gual}\ \emph {et~al.}(2016)\citenamefont
  {Sanchis-Gual}, \citenamefont {Degollado}, \citenamefont {Montero},
  \citenamefont {Font},\ and\ \citenamefont {Herdeiro}}]{Sanchis-Gual:2015lje}%
  \BibitemOpen
  \bibfield  {author} {\bibinfo {author} {\bibfnamefont {N.}~\bibnamefont
  {Sanchis-Gual}}, \bibinfo {author} {\bibfnamefont {J.~C.}\ \bibnamefont
  {Degollado}}, \bibinfo {author} {\bibfnamefont {P.~J.}\ \bibnamefont
  {Montero}}, \bibinfo {author} {\bibfnamefont {J.~A.}\ \bibnamefont {Font}}, \
  and\ \bibinfo {author} {\bibfnamefont {C.}~\bibnamefont {Herdeiro}},\ }\href
  {\doibase 10.1103/PhysRevLett.116.141101} {\bibfield  {journal} {\bibinfo
  {journal} {Phys. Rev. Lett.}\ }\textbf {\bibinfo {volume} {116}},\ \bibinfo
  {pages} {141101} (\bibinfo {year} {2016})},\ \Eprint
  {http://arxiv.org/abs/1512.05358} {arXiv:1512.05358 [gr-qc]} \BibitemShut
  {NoStop}%
\bibitem [{\citenamefont {Bosch}\ \emph {et~al.}(2016)\citenamefont {Bosch},
  \citenamefont {Green},\ and\ \citenamefont {Lehner}}]{Bosch:2016vcp}%
  \BibitemOpen
  \bibfield  {author} {\bibinfo {author} {\bibfnamefont {P.}~\bibnamefont
  {Bosch}}, \bibinfo {author} {\bibfnamefont {S.~R.}\ \bibnamefont {Green}}, \
  and\ \bibinfo {author} {\bibfnamefont {L.}~\bibnamefont {Lehner}},\ }\href
  {\doibase 10.1103/PhysRevLett.116.141102} {\bibfield  {journal} {\bibinfo
  {journal} {Phys. Rev. Lett.}\ }\textbf {\bibinfo {volume} {116}},\ \bibinfo
  {pages} {141102} (\bibinfo {year} {2016})},\ \Eprint
  {http://arxiv.org/abs/1601.01384} {arXiv:1601.01384 [gr-qc]} \BibitemShut
  {NoStop}%
\bibitem [{\citenamefont {Baake}\ and\ \citenamefont
  {Rinne}(2016)}]{Baake:2016oku}%
  \BibitemOpen
  \bibfield  {author} {\bibinfo {author} {\bibfnamefont {O.}~\bibnamefont
  {Baake}}\ and\ \bibinfo {author} {\bibfnamefont {O.}~\bibnamefont {Rinne}},\
  }\href {\doibase 10.1103/PhysRevD.94.124016} {\bibfield  {journal} {\bibinfo
  {journal} {Phys. Rev.}\ }\textbf {\bibinfo {volume} {D94}},\ \bibinfo {pages}
  {124016} (\bibinfo {year} {2016})},\ \Eprint
  {http://arxiv.org/abs/1610.08352} {arXiv:1610.08352 [gr-qc]} \BibitemShut
  {NoStop}%
\bibitem [{\citenamefont {Zilhão}\ \emph {et~al.}(2015)\citenamefont
  {Zilhão}, \citenamefont {Witek},\ and\ \citenamefont
  {Cardoso}}]{Zilhao:2015tya}%
  \BibitemOpen
  \bibfield  {author} {\bibinfo {author} {\bibfnamefont {M.}~\bibnamefont
  {Zilhão}}, \bibinfo {author} {\bibfnamefont {H.}~\bibnamefont {Witek}}, \
  and\ \bibinfo {author} {\bibfnamefont {V.}~\bibnamefont {Cardoso}},\ }\href
  {\doibase 10.1088/0264-9381/32/23/234003} {\bibfield  {journal} {\bibinfo
  {journal} {Class. Quant. Grav.}\ }\textbf {\bibinfo {volume} {32}},\ \bibinfo
  {pages} {234003} (\bibinfo {year} {2015})},\ \Eprint
  {http://arxiv.org/abs/1505.00797} {arXiv:1505.00797 [gr-qc]} \BibitemShut
  {NoStop}%
\bibitem [{\citenamefont {Herdeiro}\ \emph {et~al.}(2016)\citenamefont
  {Herdeiro}, \citenamefont {Radu},\ and\ \citenamefont
  {Runarsson}}]{Herdeiro:2016tmi}%
  \BibitemOpen
  \bibfield  {author} {\bibinfo {author} {\bibfnamefont {C.}~\bibnamefont
  {Herdeiro}}, \bibinfo {author} {\bibfnamefont {E.}~\bibnamefont {Radu}}, \
  and\ \bibinfo {author} {\bibfnamefont {H.}~\bibnamefont {Runarsson}},\ }\href
  {\doibase 10.1088/0264-9381/33/15/154001} {\bibfield  {journal} {\bibinfo
  {journal} {Class. Quant. Grav.}\ }\textbf {\bibinfo {volume} {33}},\ \bibinfo
  {pages} {154001} (\bibinfo {year} {2016})},\ \Eprint
  {http://arxiv.org/abs/1603.02687} {arXiv:1603.02687 [gr-qc]} \BibitemShut
  {NoStop}%
\bibitem [{\citenamefont {Herdeiro}\ and\ \citenamefont
  {Radu}(2017)}]{Herdeiro:2017phl}%
  \BibitemOpen
  \bibfield  {author} {\bibinfo {author} {\bibfnamefont {C.~A.~R.}\
  \bibnamefont {Herdeiro}}\ and\ \bibinfo {author} {\bibfnamefont
  {E.}~\bibnamefont {Radu}},\ }\href {\doibase 10.1103/PhysRevLett.119.261101}
  {\bibfield  {journal} {\bibinfo  {journal} {Phys. Rev. Lett.}\ }\textbf
  {\bibinfo {volume} {119}},\ \bibinfo {pages} {261101} (\bibinfo {year}
  {2017})},\ \Eprint {http://arxiv.org/abs/1706.06597} {arXiv:1706.06597
  [gr-qc]} \BibitemShut {NoStop}%
\bibitem [{\citenamefont {Ganchev}\ and\ \citenamefont
  {Santos}(2018)}]{Ganchev:2017uuo}%
  \BibitemOpen
  \bibfield  {author} {\bibinfo {author} {\bibfnamefont {B.}~\bibnamefont
  {Ganchev}}\ and\ \bibinfo {author} {\bibfnamefont {J.~E.}\ \bibnamefont
  {Santos}},\ }\href {\doibase 10.1103/PhysRevLett.120.171101} {\bibfield
  {journal} {\bibinfo  {journal} {Phys. Rev. Lett.}\ }\textbf {\bibinfo
  {volume} {120}},\ \bibinfo {pages} {171101} (\bibinfo {year} {2018})},\
  \Eprint {http://arxiv.org/abs/1711.08464} {arXiv:1711.08464 [gr-qc]}
  \BibitemShut {NoStop}%
\bibitem [{\citenamefont {Degollado}\ \emph {et~al.}(2018)\citenamefont
  {Degollado}, \citenamefont {Herdeiro},\ and\ \citenamefont
  {Radu}}]{Degollado:2018ypf}%
  \BibitemOpen
  \bibfield  {author} {\bibinfo {author} {\bibfnamefont {J.~C.}\ \bibnamefont
  {Degollado}}, \bibinfo {author} {\bibfnamefont {C.~A.~R.}\ \bibnamefont
  {Herdeiro}}, \ and\ \bibinfo {author} {\bibfnamefont {E.}~\bibnamefont
  {Radu}},\ }\href {\doibase 10.1016/j.physletb.2018.04.052} {\bibfield
  {journal} {\bibinfo  {journal} {Phys. Lett.}\ }\textbf {\bibinfo {volume}
  {B781}},\ \bibinfo {pages} {651} (\bibinfo {year} {2018})},\ \Eprint
  {http://arxiv.org/abs/1802.07266} {arXiv:1802.07266 [gr-qc]} \BibitemShut
  {NoStop}%
\bibitem [{\citenamefont {East}\ and\ \citenamefont {Pretorius}(2013)}]{best}%
  \BibitemOpen
  \bibfield  {author} {\bibinfo {author} {\bibfnamefont {W.~E.}\ \bibnamefont
  {East}}\ and\ \bibinfo {author} {\bibfnamefont {F.}~\bibnamefont
  {Pretorius}},\ }\href {\doibase 10.1103/PhysRevD.87.101502} {\bibfield
  {journal} {\bibinfo  {journal} {Phys. Rev.}\ }\textbf {\bibinfo {volume}
  {D87}},\ \bibinfo {pages} {101502} (\bibinfo {year} {2013})},\ \Eprint
  {http://arxiv.org/abs/1303.1540} {arXiv:1303.1540 [gr-qc]} \BibitemShut
  {NoStop}%
\bibitem [{\citenamefont {Yoshino}\ and\ \citenamefont
  {Kodama}(2015)}]{Yoshino:2015nsa}%
  \BibitemOpen
  \bibfield  {author} {\bibinfo {author} {\bibfnamefont {H.}~\bibnamefont
  {Yoshino}}\ and\ \bibinfo {author} {\bibfnamefont {H.}~\bibnamefont
  {Kodama}},\ }\href {\doibase 10.1088/0264-9381/32/21/214001} {\bibfield
  {journal} {\bibinfo  {journal} {Class. Quant. Grav.}\ }\textbf {\bibinfo
  {volume} {32}},\ \bibinfo {pages} {214001} (\bibinfo {year} {2015})},\
  \Eprint {http://arxiv.org/abs/1505.00714} {arXiv:1505.00714 [gr-qc]}
  \BibitemShut {NoStop}%
\bibitem [{\citenamefont {Poisson}(2004)}]{Poisson:2004cw}%
  \BibitemOpen
  \bibfield  {author} {\bibinfo {author} {\bibfnamefont {E.}~\bibnamefont
  {Poisson}},\ }\href {\doibase 10.1103/PhysRevD.70.084044} {\bibfield
  {journal} {\bibinfo  {journal} {Phys. Rev.}\ }\textbf {\bibinfo {volume}
  {D70}},\ \bibinfo {pages} {084044} (\bibinfo {year} {2004})},\ \Eprint
  {http://arxiv.org/abs/gr-qc/0407050} {arXiv:gr-qc/0407050 [gr-qc]}
  \BibitemShut {NoStop}%
\bibitem [{\citenamefont {Goncharov}\ and\ \citenamefont
  {Thrane}(2018)}]{Goncharov:2018ufi}%
  \BibitemOpen
  \bibfield  {author} {\bibinfo {author} {\bibfnamefont {B.}~\bibnamefont
  {Goncharov}}\ and\ \bibinfo {author} {\bibfnamefont {E.}~\bibnamefont
  {Thrane}},\ }\href@noop {} {\  (\bibinfo {year} {2018})},\ \Eprint
  {http://arxiv.org/abs/1805.03761} {arXiv:1805.03761 [astro-ph.IM]}
  \BibitemShut {NoStop}%
\bibitem [{\citenamefont {Pierce}\ \emph {et~al.}(2018)\citenamefont {Pierce},
  \citenamefont {Riles},\ and\ \citenamefont {Zhao}}]{Pierce:2018xmy}%
  \BibitemOpen
  \bibfield  {author} {\bibinfo {author} {\bibfnamefont {A.}~\bibnamefont
  {Pierce}}, \bibinfo {author} {\bibfnamefont {K.}~\bibnamefont {Riles}}, \
  and\ \bibinfo {author} {\bibfnamefont {Y.}~\bibnamefont {Zhao}},\ }\href@noop
  {} {\  (\bibinfo {year} {2018})},\ \Eprint {http://arxiv.org/abs/1801.10161}
  {arXiv:1801.10161 [hep-ph]} \BibitemShut {NoStop}%
\bibitem [{\citenamefont {Pretorius}(2005)}]{Pretorius:2004jg}%
  \BibitemOpen
  \bibfield  {author} {\bibinfo {author} {\bibfnamefont {F.}~\bibnamefont
  {Pretorius}},\ }\href {\doibase 10.1088/0264-9381/22/2/014} {\bibfield
  {journal} {\bibinfo  {journal} {Class. Quant. Grav.}\ }\textbf {\bibinfo
  {volume} {22}},\ \bibinfo {pages} {425} (\bibinfo {year} {2005})},\ \Eprint
  {http://arxiv.org/abs/gr-qc/0407110} {arXiv:gr-qc/0407110 [gr-qc]}
  \BibitemShut {NoStop}%
\end{thebibliography}%

\appendix
\section{Constructing initial data}
We construct initial data that satisfy the constraint part of the Einstein equations by
solving the conformal thin-sandwich equations using the methods described in~\cite{East:2013mfa}.
In this formulation, one freely chooses the conformal three-metric $\tilde{\gamma}_{ij}$,
the time derivative $\dot{\tilde{\gamma}}^{ij}$, the trace of the extrinsic curvature $K$,
and the conformal lapse $\tilde{\alpha}$. Once the Hamiltonian and momentum constraint are solved
for the conformal factor $\psi$ and the shift vector $\beta^i$, we can construct the full metric by
conformally rescaling these quantities as $\gamma_{ij}=\Psi^4\tilde{\gamma}_{ij}$, 
$\alpha=\Psi^6\tilde{\alpha}$, etc.

As indicated in the main text, we use the solutions of~\cite{East:2017ovw} describing a
Proca cloud around a spinning BH that has grown through superradiance to a
non-negligible mass, but is still far from saturation, as a starting point in
constructing initial data. These solutions have an axisymmetric spacetime, but
the complex Proca field has an $m=1$ azimuthal symmetry such that the Lie
derivative in the azimuthal direction when acting on the Proca field gives
$\mathcal{L}_{\phi}X_a=iX_a$. We set the conformal metric functions to be equal
to the values from the axisymmetric solution. In addition, we must also specify 
the energy density $U$ and three momentum density $p^i$. With the Proca field equations, there is an
additional complication since the Proca field equations also have constraints.
In order to guarantee that these will be satisfied, we make the following choices
for conformally rescaling the Proca fields.
We let $\tilde{E}^i=\psi^{6}E^i$, $\tilde{\chi}=\psi^{6}\chi$,
and $\chi_i=\tilde{\chi}_i$. 
It follows then that $B^i=\psi^{6}\tilde{B}^i$.
We set these quantities according to the Proca field
from the above mentioned complex solution, but using only the real part of the field, and rescaling
to get the same amount of energy: $X_a\rightarrow\sqrt{2}\Re{(X_a)}$. Because of this, 
the Hamiltonian and momentum constraints must be solved.

Since the conformal Proca fields satisfy the Proca field constraint
equations with respect to the conformal metric $\tilde{D}_i \tilde{E}^i=-\mu^2\tilde{\chi}$, 
it follows then that the rescaled Proca
fields will satisfy these constraints with respect to the rescaled metric:
\beq
D_iE^i= \frac{1}{\sqrt{\gamma}}\partial_i(\sqrt{\gamma}E^i) = \psi^{-6}\tilde{D}_i \tilde{E}^i=
-\mu^2\chi \ .
\eeq

The three momentum density is then given by
\beq
p_i=\epsilon_{ijk}E^jB^k+\mu^2\chi\chi_i =\psi^{-6}\tilde{p}_i
\eeq
where $\tilde{p}_i$ is the equivalent expression in terms of the conformal metric
and Proca fields.
The energy density is given by
\beq
U=\frac{1}{2}\psi^{-8}\left(\tilde{B}^i\tilde{B}_i+\tilde{E}^i\tilde{E}_i\right)
            +\frac{1}{2}\mu^2\left(\psi^{-12}\tilde{\chi}^2+\psi^{-4}\tilde{\chi}^i\tilde{\chi}_i\right) 
            \ .
\eeq
For the massless case $\mu=0$, i.e. electromagnetism, then $U=\psi^{-8}\tilde{U}$, and hence the conformal momentum
and energy have the usual dependence taken for the conformal thin-sandwich equations.
For the massive case, there is an additional contribution to the energy which does
not scale with this power, but for the cases considered here we are still able to relax
to a solution.

We also note in passing that a similar procedure can be used for constructing 
superposed initial data, for example describing a charged binary BH, without solving
the vector field constraints.
If we begin with two solutions that individually satisfy the vector field constraints,
then we can merely superpose the fields weighted by their
metric determinants, e.g.
\beq
\sqrt{\tilde{\gamma}}\tilde{E}^i = \sqrt{ {\gamma}^{(1)}}{E}^{i(1)}+\sqrt{ {\gamma}^{(2)}}{E}^{i(2)}
\eeq
to get a conformal field that when rescaled will still satisfy the constraints.

\section{Numerical convergence}
\begin{figure}
\begin{center}
\includegraphics[width=\columnwidth,draft=false]{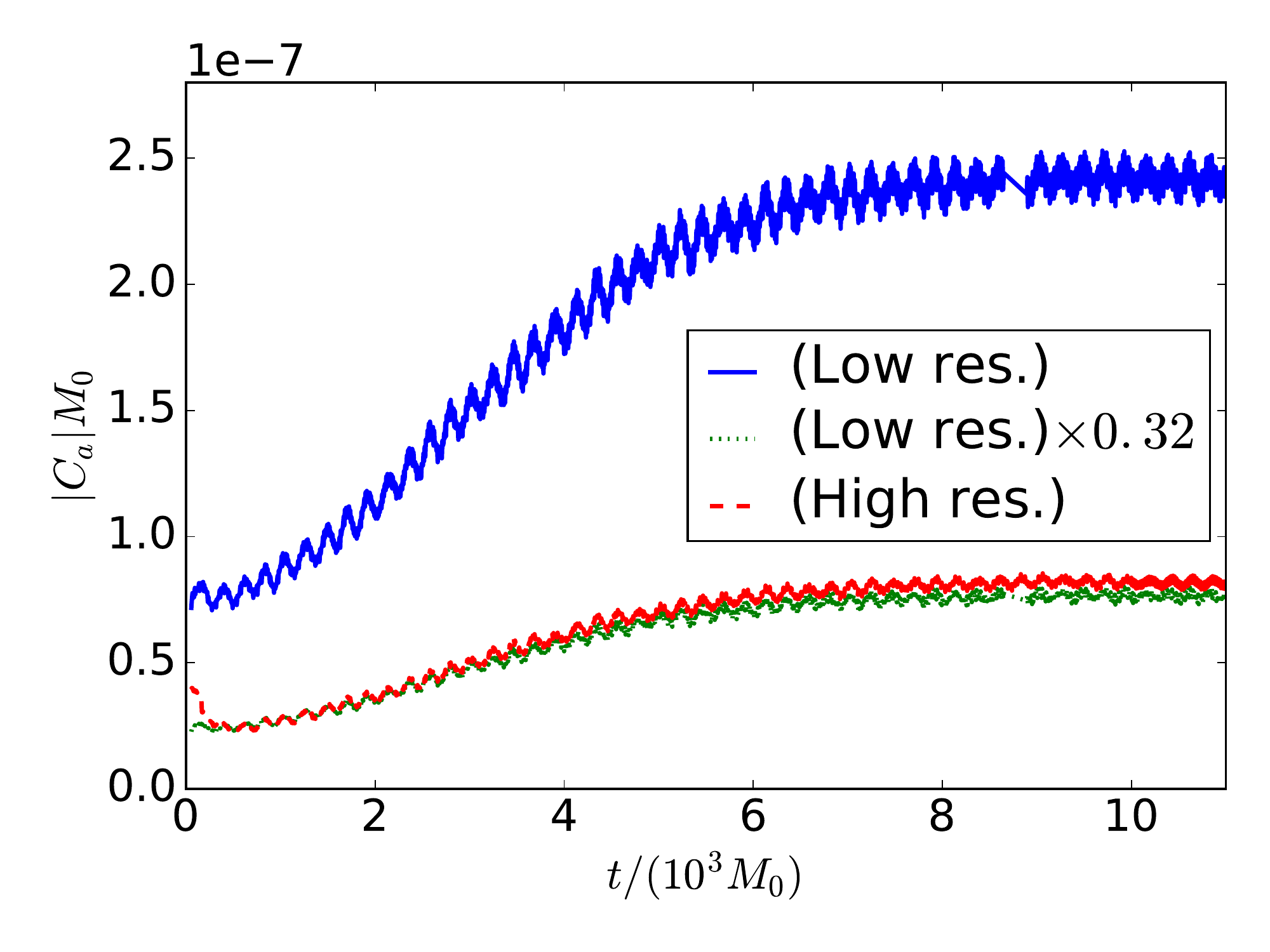}
\end{center}
\caption{
The norm of the generalized
    harmonic constraint violation $C_a=H_a-\Box x_a$ 
    (average value in the $x=0$, $(y,z)\in [-50M_0,50M_0]^2$ plane) at two resolutions.
The low resolution case is also shown rescaled by the factor expected
for fourth-order convergence.
\label{fig:cnst}
}
\end{figure}

We demonstrate that the constraints are indeed converging to zero during 
the evolution in Fig.~\ref{fig:cnst}, where we show a norm of the generalized
harmonic constraint violation $C_a=H_a-\Box x_a$~\cite{Pretorius:2004jg},
for the standard resolution and low resolution, the latter of which has $3/4\times$ the grid
spacing of the former. The decrease in this quantity with increased resolution is consistent
with the expected fourth-order convergence.

Comparing the two different resolutions, and assuming fourth-order convergence,
we can estimate the truncation error in various quantities. We estimate the
error in the saturation energy for the Proca cloud to be $\sim3\%$. For the GWs
we estimate the error in the amplitude of the dominant $\ell=m=2$ mode to be
$\sim10\%$ while the error in the smaller amplitude and higher frequency  $\ell=m=4$
mode is $\sim30\%$. For all these quantities, the trend with resolution suggests
that these quantities are underestimated due to truncation error.
\end{document}